# Observation of Nuclear Quantum Effects and Hydrogen Bond Symmetrisation in High Pressure Ice

Thomas Meier[1*], Sylvain Petitgirard[1], Saiana Khandarkhaeva[1], Leonid Dubrovinsky[1]

1) Bayerisches Geoinstitut, Bayreuth University, Universitätsstraße 30, 95447 Bayreuth, Germany

*) Thomas.meier@uni-bayreuth.de

**Abstract**
Hydrogen bond symmetrisations in H-bonded systems triggered by pressure induced nuclear quantum effects (NQEs) is a long-known concept[1] but experimental evidences in high-pressure ices have remained elusive with conventional methods[2,3]. Theoretical works predicted quantum-mechanical tunneling of protons within water ices to occur at pressures above 30 GPa and the H-bond symmetrisation transition above 60 GPa[4]. Here, we used $^1$H-NMR on high-pressure ice up to 90 GPa, and demonstrate that NQEs govern the behavior of the hydrogen bonded protons in ice VII already at significantly lower pressures than previously expected. A pronounced tunneling mode was found to be present up to the highest pressures of 90 GPa, well into the stability field of ice X, where NQEs are not anticipated in a fully symmetrized H-bond network. We found two distinct transitions in the NMR shift data at about 20 GPa and 75 GPa attributed to the step-wise symmetrization of the H-bond (HB), with high-barrier H-Bonds (HBHB)→low-barrier H-bonds (LBHB) and LBHB→ symmetric H-bonds (SHB) respectively. These transitions could have major implication on the physical properties of high-pressure ices and planetary interior models. NQEs observed in this chemically simple system over a wide pressure range could prove to be useful in designing a new generation of electronic devices exploiting protonic tunneling.

Water in its liquid and solid forms is ubiquitous in nature, being one of the most abundant molecule in the universe, and it is thought to be a prerequisite to sustain life in our solar system and beyond. Water is one the main constituent of ocean exoplanets and icy moons like Ganymede, Europa, Enceladus and Titan with possible existence of deep high-pressure ice layers in their internal structure. The hydrosphere of these bodies could be up to 900 km thick in icy satellites and up to several thousand kilometres in Ocean exoplanets[5,6]. Understanding their internal structure and evolution is crucial to determine their potential habitability and for interpreting upcoming NASA Europa Clipper and ESA Juice space missions[7,8].

Water molecules have been known for a long time to form a very specific type of chemical bonding – hydrogen bonds [9]. Under high pressure, the phase diagram of $H_2O$ exhibit an exotic behaviour with more than 15 stable crystalline phases at variable temperature and pressure conditions[10]. The high-pressure region, above 3 GPa, is mostly dominated by the three ice phases VII, VIII and X (figure 1). Both ice VII and VIII are molecular solids consisting of distinct $H_2O$ units linked to each other by hydrogen bonds. Ice X, on the other hand, exhibits a fully symmetrised hydrogen bond network, thus rendering a dissociation of the $H_2O$ molecules, forming an atomic solid at pressures of about 50 to 70 GPa at room temperature[4]. One of the most enigmatic phenomena in the high-pressure phase diagram of water is the transition from the hydrogen disordered phase ice VII into the hydrogen ordered phase of ice X [11]. It is widely believed that this transition is preceded by nuclear quantum effects (NQEs)[4], or specifically, pronounced proton delocalisation due to tunnelling motion within the symmetric double-well potential of the hydrogen bonds in ice VII[17].

Evidence of hydrogen-bond symmetrisation and potential room-temperature proton tunnelling in ice VII are sparse and often contradictory. This obviously relates to experimental difficulties. Hydrogen atoms remain effectively invisible to X-ray diffraction or emission spectroscopy[12], leaving for observations only the heavier oxygen sublattice, which does not show significant transitions in the pressure range of interest[13,14]. While Raman-spectroscopy and neutron diffraction are more sensitive to H-bonds, they also yield ambiguous and often contradictory results[15,16].

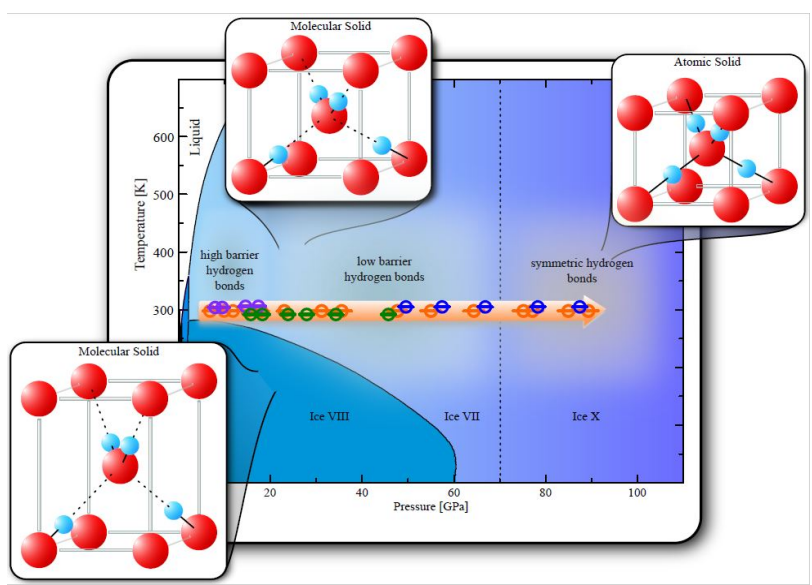

*Figure 1: Phase diagram of $H_2O$. The insets show the schematic symmetries of the high pressure phases of ice VII, VIII and X. Round symbols denote NMR measurements within this study, different colors relate to several pressure runs and DACs.*

Given these experimental difficulties, a direct observation of H-bond symmetrisation or NQEs remain mostly elusive in high-pressure experiments with the techniques discussed above.

One of the most promising spectroscopic methods to deal with these problems is Nuclear Magnetic Resonance (NMR) spectroscopy, where proton NMR is best known for providing one of the highest possible NMR signal strengths, and allows direct observation of the electronic and structural environment of the nuclei. However, an application of NMR spectroscopy at high pressures, particularly in diamond anvil cells (DACs), was unfeasible, with set-ups unable to surpass 8 GPa on average, with rare exceptions reaching pressures above 10 GPa [17].

In this study, we used a novel NMR resonator structure[18] (see figure 2A) at pressures approaching the megabar regime to investigate compression-induced nuclear quantum effects in ices. With this novel approach, we were able to follow the hydrogen bond symmetrization in ice VII and its transition to the proton ordered phase X, one of the most sought after elusive effects in high pressure sciences proposed 46 years ago.

Our $^1$H-NMR spectra on ice from 8 to 90 GPa are shown in figure 2B. We performed several experiments using four independently loaded cells (see Methods), with overlapping and reproducible results (Fig 2D).

NQEs of protons have several observable effects on $^1$H-NMR spectra, and can be interpreted based on methyl groups [19–21] studies at ambient pressures and cryogenic temperatures. Firstly, random rapid tunnelling from one to another minimum in a symmetric double-well energy potential will result in motional averaging of the NMR signals[22]. Secondly, the formation of a symmetric double-well potential leads to a degeneration of the nuclear spin ground state. It could be shown that this degeneration is lifted[23], leading to an observable zero-field splitting of the NMR signal, which is directly proportional to the particle's tunnel frequency[24,25]. Furthermore, ¹H-NMR spectra in low barrier hydrogen bonds (LBHBs) exhibit significant de-shielding, with high proton shifts of about 20 to 40 ppm [26].

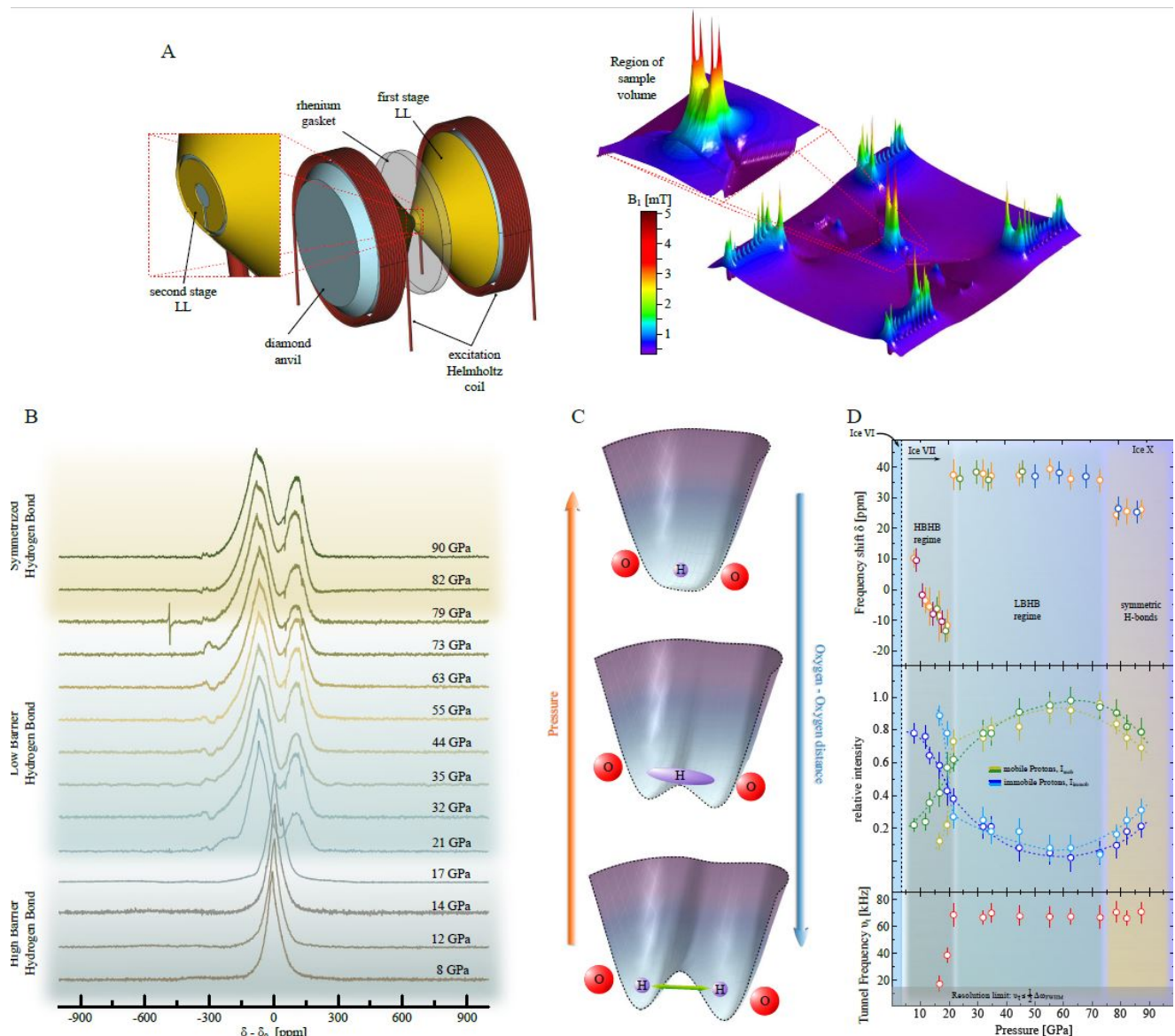

*Figure 2: A: High pressure NMR resonator set-up. The inset shows the zoomed-in region around the anvil's culet. Both Lenz lenses are formed on the anvil's pavilion by copper deposition and subsequent shaping, using a focused ion beam(left). Simulations of the RF magnetic field generated by the resonator set-up demonstrate the magnification in $B_1$ at the sample cavity (right). B: Proton NMR signals of ice from 8 to 90 GPa. C: Sequence of hydrogen bond symmetrization with pressure. D: Upper panel: chemical shifts relative to a water sample at ambient conditions. Middle panel: Intensities from localized and tunneling protons. Lower panel: Tunnel frequencies up to 90 GPa.*

The ¹H-NMR spectra shown in figure 2B were de-convoluted (see suppl. material) into Gaussian and Lorentzian contributions and attributed to localized immobile protons and randomly tunnelling protons delocalized over the energy hypersurface of the hydrogen bond. At pressures above 20 GPa, a pronounced tunnel splitting could be observed with tunnel frequencies between 70 and 80 kHz (figure 2D lower panel).

The integrated signal intensities associated to tunnelling ($I_{mob}$) and localised ($I_{immob}$) protons can be seen in the mid panel of figure 2D. The observed evolution of both $I_{immob}$ and $I_{mob}$ is consistent with the predicted evolution[27] of the energy barrier of the double-well potential of the H-bond (see figure 2C), i.e. increasing pressure reduces height and width of the barrier, increasing the probability of the protons to tunnel between the energy minima of the double-well potential. Thus, the relative intensity of signal due to localized protons declines, whereas the intensity of signal originating from quantum mechanical tunneling increases with decreasing oxygen-oxygen distances. At about 60 GPa, the majority of the protons participate in collective tunneling motion as $I_{mob}$ reaches its maximum. At pressures above 60 – 70 GPa, the tunnel probability declines as height and width of the energy barrier approach zero, thus localizing the protons with pressure, leading to an increase in $I_{immob}$.

The upper panel of figure 2D shows chemical shift values δ as a function of pressure. Two distinct transitions are evident at pressures of 20 and 75 GPa respectively. While the pressure of the second transition is in very good agreement with the proposed transition for a symmetric hydrogen bond network in ice X, the first transition has not been observed with other methods and can be associated with a transition from the HBHB regime to the LBHB regime[28]. Thus, our NMR shift data indicates that a "Low-Barrier Hydrogen Bond" exists in ice VII not only at significantly lower pressures, but also in a pressure range where NQEs should be absent.

Our study indicates a much more complex scenario of the interplay between pressure induced NQEs and the hydrogen bond symmetrization in high pressure ices than what could be anticipated from other experimental work in this field. In fact, up to this point no clear experimental distinction between the HBHB and LBHB regimes could be defined, whereas theoretical estimates vary often by tens of GPa between 30 and 60 GPa. Moreover, it could be shown that the LBHB -> SHB transition is indeed not a continuous one but exhibits a clear transition pressure of about 75 GPa. Such behavior in high-pressure ice may have direct influences on their mechanical properties or capabilities to incorporate salt in their structure[29] with implication for planetary interior modeling.

Our study demonstrates the possibility to directly trigger NQEs at room temperature solely by the application of relatively low pressures, for one of the chemically simplest substances - water. Therein, it closely resembles observed effects of pressure driven superconductivity[30], and might well lead to an unprecedented surge for pressure tuned quantum devices, employing heavier quantum-mechanical particles than electrons exhibiting room temperature NQEs. The advantage of such proton tunneling devices would be their significantly reduced size as NQEs can be triggered within the length scale of a hydrogen bond, i.e. about 2.3 to 2.5 Å, which is often much below the dimensions of common electronic devices employing electronic diffusion like MOSFETs or tunneling like TFETs. Furthermore, this work shows that a wide class of materials could exhibit such pressure induced NQEs as hydrogen bonds are ubiquitous in a wide range of chemical compounds.

*Author Contributions*

T.M. designed and built the NMR resonator, prepared DACs and conducted experiments. S.P. and S.K. performed the FIB based shaping of the Lenz lenses. T.M., S.P. and L.D. analysed the data and wrote the manuscript.

*Acknowledgements*

We would like to thank Professor Ernst Rössler and Thomas Körber for provision of the 9.3 T NMR system. Furthermore, we thank Nobuyoshi Miyajima and Katharina Marquardt for provision of the FIB, and help with the ion milling (grant number: INST 90/315-1 FUGG). The authors, T.M. and L.D., were funded by the Bavarian Geoinstitute through the Free State of Bavaria. S.P. and S.K. were financed by the german research society (PE 2334/1-1 and DU-393/13-1).